\documentclass[aps,floatfix,twocolumn,showpacs,amsmath,amssymb,superscriptaddress,longbibliography]{revtex4-1} 

\usepackage{graphicx}
\usepackage{dcolumn}
\usepackage{color}
\usepackage{amsmath}
\usepackage[breaklinks,colorlinks,bookmarks=false,citecolor=blue,linkcolor=red,urlcolor=blue]{hyperref}
\newcommand{\ve}[1]{\boldsymbol{#1}}

\def\abs#1{\left|#1\right|}

\begin{document}

\title{The Hubbard model on the honeycomb lattice: from static and dynamical mean-field theories to lattice quantum Monte Carlo simulations}

\author{Marcin~Raczkowski}
\affiliation{Institut f\"ur Theoretische Physik und Astrophysik,
             Universit\"at W\"urzburg, Am Hubland, D-97074 W\"urzburg, Germany}

\author{Robert~Peters}
\affiliation{Department of Physics, Kyoto University, Kyoto 606-8502, Japan}

\author{Th\d{i}~Thu~Ph\`ung}
\affiliation{Laboratoire de Physique Th\'eorique et Mod\'elisation, CNRS UMR 8089,
  CY Cergy Paris Universit\'e, F-95302 Cergy-Pontoise Cedex, France}
\affiliation{University of Science and Technology of HaNoi, 18 Hoang Quoc Viet, Vietnam}

\author{Nayuta~Takemori}
\affiliation{Research Institute for Interdisciplinary Science, Okayama University, Okayama, 700-8530, Japan}

\author{Fakher~F.~Assaad}
\affiliation{Institut f\"ur Theoretische Physik und Astrophysik,
             Universit\"at W\"urzburg, Am Hubland, D-97074 W\"urzburg, Germany}

\author{Andreas~Honecker}
\affiliation{Laboratoire de Physique Th\'eorique et Mod\'elisation, CNRS UMR 8089,
  CY Cergy Paris Universit\'e, F-95302 Cergy-Pontoise Cedex, France}

\author{Javad~Vahedi}
\affiliation{Laboratoire de Physique Th\'eorique et Mod\'elisation, CNRS UMR 8089,
  CY Cergy Paris Universit\'e, F-95302 Cergy-Pontoise Cedex, France}
\affiliation{Department of Physics, Sari Branch, Islamic Azad University, Sari 48164-194, Iran}

\date{\today}

\begin{abstract}
We study the one-band Hubbard model on the honeycomb lattice using a 
combination of quantum Monte Carlo (QMC) simulations and static as well as 
dynamical mean-field theory (DMFT). This model is known to show a quantum 
phase transition between a Dirac semi-metal and the antiferromagnetic 
insulator. The aim of this article is to provide a detailed comparison 
between these approaches by computing static properties, notably 
ground-state energy, single-particle gap, double occupancy, and staggered 
magnetization, as well as dynamical quantities such as the single-particle 
spectral function. At the static mean-field level local moments cannot be 
generated without breaking the SU(2) spin symmetry. The DMFT approximation 
accounts for temporal fluctuations, thus captures both the evolution of the 
double occupancy and the resulting local moment formation in the 
paramagnetic phase. As a consequence, the DMFT approximation is found to 
be very accurate in the Dirac semi-metallic phase where local moment 
formation is present and the spin correlation length small. However, in 
the vicinity of the fermion quantum critical point the spin correlation 
length diverges and the spontaneous SU(2) symmetry breaking leads to 
low-lying Goldstone modes in the magnetically ordered phase. The impact of 
these spin fluctuations on the single-particle spectral function -- 
\textit{waterfall} features and narrow spin-polaron bands -- is only 
visible in the lattice QMC approach.
\end{abstract}

\maketitle

\section{Introduction}
\label{sec:intro}

The one-band ``Hubbard'' model 
\cite{Hubbard1963,Kanamori1963,Gutzwiller1963,Tasaki1998,Eder2017} is one 
of the basic models for correlation effects in solids. Its square-lattice 
version has been investigated extensively because of its relevance to the 
high-temperature superconductors \cite{Dagotto1994,Brenig1995}. Screened 
electronic correlations modeled by a Hubbard-$U$ term generate local 
magnetic moments.
For the half-filled band,
local moments generically order and the global SU(2) spin 
symmetry is spontaneously broken leading to Goldstone modes.
The interplay of charge and 
spin degrees of freedom is the key point captured by the Hubbard and 
strong-coupling $t$-$J$ models. For the well-studied single-hole problem,
the single-particle spectral function of 
the square-lattice Hubbard and $t$-$J$ models reveals spin polaron 
quasiparticles as well as ``waterfall'' features 
\cite{Preuss95,Preuss1997,Brunner00b}. These anomalous spectral properties 
and their evolution with doping have been the subject of extensive 
numerical studies 
\cite{PRB.73.165114,Macridin2007,Zemljic2008,PRB.78.064501, 
Moritz10,PRB.82.134505,Piazza2012,Rost2012,PRB.90.035111,Yang2016,PRB.97.115120}.

For the half-filled Hubbard model on the square lattice, perfect nesting
drives the system into an antiferromagnetic phase for any finite on-site repulsion $U>0$
\cite{Hirsch85,White89}. By contrast, while the honeycomb lattice is also
bipartite, the half-filled Hubbard model on this lattice is distinguished
by a vanishing density of states at the Fermi level such that a finite
$U$ is required to drive the system into the antiferromagnetic phase that is
expected for large $U$,
as was already remarked in the seminal work Ref.~\cite{Sorella1992}. On the one
hand, having access to a transition from a Dirac semi-metal to an antiferromagnet
at a finite value of $U$ is interesting from a fundamental point of
view since it allows one, e.g., to study the critical properties. On the
other hand, graphene \cite{Geim2007,RevModPhys81,Yazyev2010,Wakabayashi13}
is believed to be well described by the Hubbard model on the honeycomb 
lattice in its semi-metallic phase such that weak-coupling methods remain
appropriate tools.

Extensive numerical studies of the phase diagram of the half-filled Hubbard
model on the honeycomb lattice
\cite{Sorella1992,nature2012,Sorella2012,He2012,Hassan2013,Seki2013,Assaad2013,Wu2014}
have led to the consensus that the transition between the 
paramagnetic semi-metal and the antiferromagnetic insulator is a direct 
one with an unusual quantum critical point separating these two phases.
The critical behavior is captured by a Gross-Neveu-Yukawa field theory \cite{Herbut09a},
consisting of eight-component Dirac fermions \cite{Ryu09}
as well as a three-component $\phi^{4}$-theory accounting for the 
magnetic order parameter and low-lying long-wave-length Goldstone modes. 
The Yukawa term couples the three-component bosonic modes to the triplet 
of antiferromagnetic mass terms such that when the bosons condense fermion 
mass is generated. The upper critical dimension for this theory is of 
three spatial dimensions such that an $\epsilon$-expansion can be used to
calculate the deviation of the critical exponents from the mean-field
results in two dimensions \cite{Herbut09a}.

The aim of this paper is to provide a detailed comparison between various 
approximations and numerically exact quantum Monte Carlo (QMC) results for 
the Hubbard model on the honeycomb lattice. We will start with the 
mean-field approximation that is widely used in the context of graphene,
see  Refs.~\cite{Yazyev2010,Wakabayashi13,Feldner2010,*FeldnerE} and references therein. 
The first clear shortcoming of this 
approximation is the failure to generate local moments without breaking 
the SU(2) spin symmetry. The minimal extension of the static mean-field 
approximation to account for local moment formation is dynamical 
mean-field theory (DMFT). Provided that the magnetic correlation length is 
not too big, DMFT is expected to provide a good account of the 
physics, and thus promises improved numerical accuracy in the parameter
regime relevant to graphene at a moderate computational cost. 

Early single-site DMFT studies located the metal-insulator transition 
around $U_c/t \gtrsim 10$ \cite{Jafari2009,Tran2009}. This is not only 
significantly above the mean-field transition $U_c/t \approx 2.23$
and the early QMC estimate $U_c/t = 4.5 \pm 0.5$
\cite{Sorella1992}, but also much larger than the most accurate QMC 
results, namely $U_c/t \approx 3.87$ \cite{Sorella2012} and $U_c/t \approx 
3.78$ \cite{Assaad2013}, respectively. Consequently, further 
investigations of the semi-metal--antiferromagnet transition in the 
Hubbard model on the honeycomb lattice focused on cluster and other 
extensions of DMFT 
\cite{He2012,Hassan2013,Seki2013,Liebsch2013,Wu2014,Hirschmeier2018}, and 
the corresponding estimates for the location of the transition 
converge to the region $U_c/t \approx 3.6 \ldots 3.8$ 
\cite{Wu2014,Hirschmeier2018}, see Ref.~\cite{Hirschmeier2018} for a more 
detailed summary. These estimates from different generalizations of DMFT 
are indeed very close to the QMC estimates. However, the single-site 
studies \cite{Jafari2009,Tran2009} only looked at paramagnetic solutions. 
Thus, to the best of our knowledge, the accuracy of the simple single-site 
DMFT when one allows for the relevant antiferromagnetic solution at large 
$U$ has not been investigated in the literature. Hence, we implement this 
here and benchmark it against QMC results on the lattice.

Furthermore, the spectral functions of the Hubbard model on an infinite 
honeycomb lattice are in principle well known, at least at the mean-field 
level, but to the best of our knowledge they have not been explicitly 
shown in the literature. Hence, we will discuss mean-field results here, 
and compare them to more elaborate DMFT and lattice QMC results.
Among others, we will show that both the spin-polaron physics and the
so-called \textit{waterfall} features known from the square lattice
are also present close to the quantum critical point on the honeycomb
lattice, but that QMC simulations are required to reveal them.

The outline of the paper is as follows: In Sec.~\ref{sec:Methods} we 
introduce the model and the three methods that we employ for our 
comparative discussion. Section \ref{eq:secInfStatic} focuses on static 
properties and the dynamical properties are investigated via spectral 
functions in Sec.~\ref{sec:SpecFns}. We summarize our findings and provide 
perspectives in Sec.~\ref{sec:Conclusions}.

\section{Model and methods}

\label{sec:Methods}

\begin{figure}[t!]
\centering
\includegraphics[width=0.63\columnwidth]{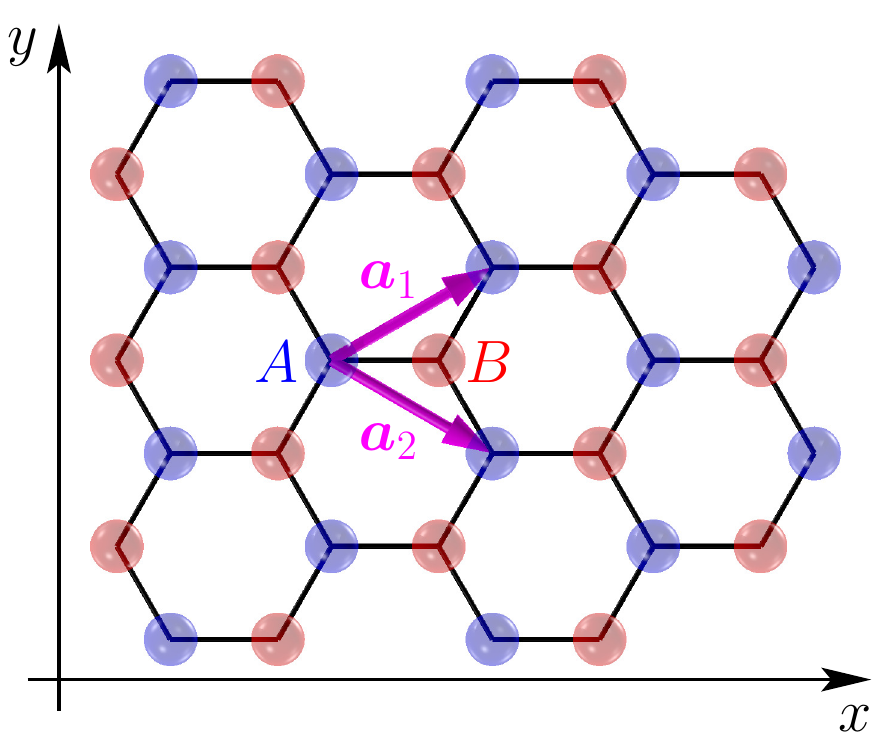}
\caption{Sketch of the honeycomb lattice showing the two sites
$A$ and $B$ in the primitive cell and our choice of primitive vectors
$\ve{a}_1$, $\ve{a}_2$.
\label{fig:honeycomb_lattice}
}
\end{figure}

We study the Hubbard model whose Hamiltonian reads
\begin{equation}
H=-t\sum_{\langle i,j\rangle \atop \sigma=\uparrow,\downarrow}c^{\dagger}_{i,\sigma}c_{j,\sigma}
+ U\, \sum_i \left(n_{i,\uparrow}-\frac{1}{2}\right)\,\left(n_{i,\downarrow} -\frac{1}{2}\right)
\label{eq:Hubbard}
\end{equation}
with $n_{i,\sigma}=c^{\dagger}_{i,\sigma}c_{i,\sigma}$. Here $\langle 
i,j\rangle $ are nearest neighbors on a lattice that we take to be 
the honeycomb lattice illustrated in Fig.~\ref{fig:honeycomb_lattice}. We 
will be interested either in the infinite system, or in a finite but large 
one. In the latter case, we denote the total number of sites by $N$ and 
impose periodic boundary conditions.

Note that since the honeycomb lattice is bipartite, the single-band 
Hubbard model on this lattice is particle-hole symmetric (see, for 
example, Ref.~\cite{Scalettar16}), {\it i.e.}, upon exchanging electron 
creation and annihilation operators, one finds a Hamiltonian that is 
equivalent to the original one of Eq.~(\ref{eq:Hubbard}). This 
particle-hole symmetry ensures that the global ground state is found at 
half filling, {\it i.e.}, for an average of one electron per lattice site.

\subsection{Static mean-field theory (MFT)}

Many authors have used a real-space Hartree-Fock-type mean-field 
approximation to study magnetism in graphene, see 
Refs.~\cite{Yazyev2010,Wakabayashi13,Feldner2010,*FeldnerE} and references therein. 
Here, we exploit the SU(2)-symmetry of the original Hubbard model 
Eq.~(\ref{eq:Hubbard}) to align the quantization axis with a possible 
ordered moment. Then the Hartree-Fock approximation amounts to
\begin{eqnarray}
H^{\rm MF}&=& -t\sum_{\langle i,j\rangle \atop \sigma=\uparrow,\downarrow}c^{\dagger}_{i,\sigma}c_{j,\sigma}
\label{eq:HMF1}\\
&&
+U\,\sum_i \Bigl( \langle n_{i,\uparrow} \rangle n_{i,\downarrow} 
 + n_{i,\uparrow}\langle n_{i,\downarrow}\rangle
- \langle n_{i,\uparrow}\rangle \langle n_{i,\downarrow}\rangle \quad
\label{eq:HMF2} \\
&& \qquad\quad -\frac{n_{i,\uparrow}+ n_{i,\downarrow}}{2}
+ \frac{1}{4} \Bigr) \, .
\label{eq:HMF3}
\end{eqnarray}
Note that the last term in Eq.~(\ref{eq:HMF2}) could be omitted for most 
purposes, but it is needed if one wants to compare total energies with the 
original Hubbard model Eq.~(\ref{eq:Hubbard}). The density-dependent term 
in Eq.~(\ref{eq:HMF3}) ensures half filling in the grand-canonical 
description thanks to particle-hole symmetry.

Although we have formulated the problem above in real space, here we will 
actually work in reciprocal space because we are interested in an infinite 
system. Since the primitive cell contains two sites, we will need to 
diagonalize a $2 \times 2$ matrix for each value of the momentum $\ve{k}$, 
followed by a summation over $\ve{k}$.

To be specific, we first perform a Fourier transformation
\begin{equation}
c_{\ve{r},\alpha,\sigma}
 = \sqrt{\frac{2}{N}}\,\sum_{\ve{k}} {\rm e}^{i\,\ve{k}\cdot \ve{r}} c_{\alpha,\sigma}(\ve{k}) \, ,
\end{equation}
where $\alpha = A,B$ labels the two sites in the primitive cell,
$\ve{r}$ is the real-space position of the primitive cell,
and $N/2$ corresponds to the number of primitive cells.
We further restrict to half filling and express all the densities
in terms of the staggered magnetization $m_{\rm s}$
\begin{equation}
\langle n_{A,\sigma}\rangle = \frac12 + \sigma\, m_{\rm s} \, , \quad
\langle n_{B,\sigma}\rangle = \frac12 - \sigma\, m_{\rm s} \, ,
\label{eq:MFms}
\end{equation}
where we wrote $\sigma = +1$ for the spin up configuration ($\uparrow$) and
$\sigma = -1$ for the spin down configuration ($\downarrow$).
With these notations and dropping the ``constant'' term Eq. (\ref{eq:HMF3}),
the mean-field Hamiltonian of Eqs.~(\ref{eq:HMF1}), (\ref{eq:HMF2})
can now be cast in the form
\begin{widetext}
\begin{equation}
 H^{\rm MF} = \sum_{\ve{k},\sigma} \left(c^\dagger_{A,\sigma}(\ve{k}), c^\dagger_{B,\sigma}(\ve{k})\right)\,
 \left( \tau_x \, \text{Re}z(\ve{k}) + \tau_y \, \text{Im}z(\ve{k}) - \ U\,\sigma \,m_{\rm s}\, \tau_z \right)
\, \begin{pmatrix}
c_{A,\sigma}(\ve{k}) \\ c_{B,\sigma}(\ve{k})
\label{eq:HMFk}
\end{pmatrix} \, ,
\end{equation}
\end{widetext}
where the Pauli matrices $\tau_{x,y,z} $ act on the ``orbital'' index $A$, $B$
and
$z(\ve{k}) = -t \left( 1 + {\rm e}^{-i \ve{k} \cdot \ve{a}_1} + {\rm e}^{-i \ve{k} \cdot \ve{a}_2} \right)$.
Here the primitive vectors are
$\ve{a}_1 = a\left( \frac{\sqrt{3}}{2}, \frac{1}{2} \right)$,
$\ve{a}_2 = a\left( \frac{\sqrt{3}}{2}, - \frac{1}{2} \right)$,
compare Fig.~\ref{fig:honeycomb_lattice}, and the lattice constant
of the underlying triangular lattice is denoted by $a$.

{}From the $2 \times 2$ matrix in Eq.~(\ref{eq:HMFk}) one immediately gets
the single-particle dispersion
\begin{equation}
E_{\pm}(\ve{k}) = \pm E(\ve{k})
\text{ with }
E(\ve{k}) = \sqrt{\abs{z(\ve{k})}^2 + \left(U\,m_{\rm s} \right)^2} \,.
\label{eq:MFTdisp}
\end{equation} 
At the Dirac points $\mathbf{K}$ (see inset of Fig.~\ref{fig:AkomegaMFT}(c)
below for a definition), we have $z(\mathbf{K}) = 0$ such that we find the
single-particle gap
\begin{equation}
\Delta_{\rm sp} = E( \mathbf{K}) = U\,\abs{m_{\rm s}}\, .
\label{eq:MFTspGap}
\end{equation} 
Thus, a finite staggered magnetization leads to the opening of a mass gap
in the spectrum.

The staggered magnetization, $m_{\rm s}$, still needs to be determined
self-consistently such that Eq.~(\ref{eq:MFms}) holds. In the following
sections, we will use a numerical solution that has been obtained by
iteration, {\it i.e.}, starting with a guess for $m_{\rm s}$, then
recomputing it via Eq.~(\ref{eq:MFms}) until convergence is reached.

On the other hand, we can make analytic progress by considering only 
low-energy physics. First, we cast the self-consistency condition for 
$m_{\rm s}$ in the gap equation
\begin{equation}
 1 = 2 U \int {\rm d} \epsilon N(\epsilon)
 \frac{1}{\sqrt{\epsilon^2 + \Delta_{\rm sp}^2}}
 \tanh \left( \frac{1}{2\,T} \sqrt{\epsilon^2 + \Delta_{\rm sp}^2}\right) 
\label{eq:gapEq}
\end{equation}
with density of states
\begin{equation}
 N(\epsilon) = \frac{2}{N} \sum_{\ve{k}}
  \delta \left(\abs{z(\ve{k})} - \epsilon \right) \, .
\end{equation}
Linearizing around the Dirac points allows for an analytic solution. Let 
\begin{equation}
	 \abs{z(\mathbf{K} + \ve{p})} \simeq v_F\, \abs{\ve{p}}   
\end{equation}
such that 
\begin{equation}
	N(\epsilon)  = \frac{\epsilon} {\pi v_F^2} \Theta(\epsilon)
\end{equation}
with $\Theta(\epsilon)$ the Heaviside function. 
Next, we introduce a high-energy cutoff $\Lambda $ to ensure
that $ \int_{0}^{\Lambda} {\rm d} \epsilon N(\epsilon) = 1$.
This yields $ \Lambda = \sqrt{2\pi} v_F$ and at zero temperature 
the gap equation Eq.~(\ref{eq:gapEq}) reduces to 
\begin{equation}
 1 = \frac{2U}{\pi v_F^2} \left( \sqrt{\Delta_{\rm sp}^2 + \Lambda^2} - \Delta_{\rm sp} \right) \, . 
\end{equation}
At $U_c$ the single-particle gap vanishes such that: 
\begin{equation}
  U_c = \frac{\sqrt{\pi}}{2 \sqrt{2}} v_F.
\end{equation}
The finite value of $U_c$ even in the presence of nesting follows from the vanishing of the density of states that cuts off the singularity in the gap equation at $\epsilon=0$
and $\Delta_{\rm sp} = 0$. For $U > U_c$ 
\begin{equation}
 m_{\rm s} = \frac{\sqrt{\pi} v_F}{\sqrt{2}U} \left( \frac{U}{U_c} - \frac{U_c}{U} \right) 
\end{equation}
such that in the vicinity of the critical point: 
\begin{equation}
 m_{\rm s} \propto \left(   U - U_c \right)^{\beta}
\end{equation}
with order parameter exponent $\beta = 1$. This mean-field value of the 
exponent stands at odds with the generic Ginzburg-Landau result $\beta =1/2$,
and demonstrates that the fermionic degrees of freedom cannot be omitted.

\subsection{Dynamical mean-field theory (DMFT)}

DMFT maps the original lattice problem onto a self-consistent quantum-impurity problem \cite{DMFT96}, which becomes exact in the limit of infinite dimension. 
This mapping is performed by calculating the local lattice Green's functions
of all atoms inside the primitive cell,
\begin{equation}
G_{i\sigma}(z)=\int {\rm d}k
\, \left(z\mathbb{I}-H_0(\ve{k})-\mathbf{\Sigma}_\sigma(z)\right)^{-1}_{ii},
\label{eq:local_Green}
\end{equation}
where $\mathbb{I}$ is the unit matrix, $H_0(\ve{k})$ the one-particle part of the Hamiltonian depending on the momentum $\ve{k}$, $\mathbf{\Sigma}_\sigma(z)$ the self-energy matrix for spin direction $\sigma=\{\uparrow,\downarrow\}$, and $i$ the index enumerating the atoms in the primitive cell.
For the honeycomb lattice, we use a primitive cell including two atoms.
By calculating the local Green's functions, DMFT takes the structure of the lattice into account. The matrix $\mathbf{\Sigma}_\sigma(z)$ includes only local self
energies; non-local parts of the self energy, e.g., a self energy between different atoms in the primitive cell, are neglected in this approach.
As discussed in the Appendix,
the latter approximation has significant consequences for the symmetry of the self energy of
a magnetic solution. Alternatively, one could use a primitive cell consisting of two sites
as a basic unit. However, such a choice would not only break the rotational symmetry of the lattice
and thus would require a treatment beyond DMFT (see, e.g., Ref.~\cite{Hirschmeier2018}), but it would also
be computationally more demanding. Therefore we decided to explore the accuracy of the simplest
single-site DMFT approximation.

By comparing between the Green's function of an Anderson impurity model and the local Green's function,
$G_{i\sigma}(z)$, in the single-site approximation, we define the hybridization function $\Delta_{i\sigma}(z)$ as
\begin{eqnarray}
G_{i\sigma}(z)=G_{\rm imp}(z)&=&\frac{1}{z-\Delta_{i\sigma}(z)-\Sigma_{i\sigma}(z)}\\
\Rightarrow \Delta_{i\sigma}(z)&=&z-\Sigma_{i\sigma}(z)-G_{i\sigma}^{-1}.
\end{eqnarray}
The hybridization function $\Delta_{i\sigma}(z)$ completely defines the coupling of an Anderson impurity to a bath of conduction electrons. 
Thus, the hybridization functions $\Delta_{1\sigma}$ and $\Delta_{2\sigma}$ together with the local two-particle interaction part of the Hamiltonian, Eq. (\ref{eq:Hubbard}), define two independent Anderson impurity models.
We note that the self energy $\Sigma_{i\sigma}$ and the hybridization function $\Delta_{i\sigma}$ depend on the spin direction. This will be important when describing magnetic states, in which $\Sigma_{i\uparrow}\neq\Sigma_{i\downarrow}$. 
 
We are using the numerical renormalization group (NRG) \cite{Wilson1975,BCT08} and
continuous-time QMC (CTHYB) \cite{Werner2006,RevModPhys.83.349,Hafermann2013} in order to
solve these resulting effective quantum impurity problems and calculate the self
energy $\Sigma_{i\sigma}(z)$.
For CTHYB, we employ the hybridization expansion CT-QMC code of the
ALPS libraries~\cite{Bauer_2011}.
The impurity self energies are used to calculate new local Green's functions,
Eq.~(\ref{eq:local_Green}). This DMFT self-consistency cycle is repeated until convergence is achieved.

Among the two different numerical techniques to solve the DMFT impurity 
problem, NRG uses a logarithmic discretization of the conduction band, 
mapping it onto a one-dimensional chain, that is iteratively diagonalized 
by discarding high-energy states \cite{Wilson1975,BCT08}. On the one hand, 
this logarithmic discretization makes it possible to calculate properties 
at $T=0$, and spectral functions for real frequencies with high accuracy 
around the Fermi energy \cite{PhysRevB.74.245114}. On the other hand, this 
logarithmic discretization leads to low accuracy in the spectral functions 
for frequencies away from the Fermi energy. Furthermore, a broadening 
function must be used to obtain a smooth Green's function and self 
energies away from the Fermi energy.

By contrast, CTHYB samples Feynman diagrams using imaginary-time Green's 
functions at finite temperature. Thus, while CTHYB can be expected to 
yield accurate results at finite temperatures for static quantities, CTHYB 
cannot directly calculate properties at $T=0$ and would require an 
analytic continuation to obtain Green's functions and self energies for 
real frequencies.

\subsection{Lattice quantum Monte Carlo (QMC)}

We have used a standard implementation of the projective auxiliary field QMC algorithm \cite{Sugiyama86,Sorella89,Imada89}. This approach is based on the equation
\begin{equation}
	\frac{ \langle \psi_0 | O | \psi_0 \rangle}{\langle \psi_0 | \psi_0 \rangle} = \lim_{\theta \rightarrow \infty} 
	\frac{ \langle \psi_T | e^{-\theta H} O e^{-\theta H}| \psi_T \rangle}{\langle \psi_T | e^{-2\theta H} | \psi_T \rangle}.
\end{equation}
Here $|\psi_0 \rangle$ is the ground state of the Hamiltonian $H$ and the 
equality holds provided that the trial wave function $|\psi_T\rangle$ is 
not orthogonal to the ground state. For practical purposes we have chosen 
the trial wave function to be the ground state of the non-interacting 
Hamiltonian. For periodic boundary conditions and lattice sizes $L=3n$ 
with integer $n$, this ground state is degenerate so that we included an 
\textit{infinitesimal} twist in the boundary condition to lift the 
degeneracy and select a ground state. For this choice of the trial wave 
function, a projection parameter $\theta = 10$ suffices to obtain 
ground-state properties on lattices with $L\times L$ unit cells up to 
$L=18$ (the number of lattice sites is $N=2\,L^2$). We have used an 
imaginary time step $\Delta \tau = 0.1$ and a symmetric Trotter 
decomposition to guarantee hermiticity of the imaginary time propagator. 
The systematic error associated with this choice of the imaginary time 
step is very small: at $U/t=6$, where it is the largest, it amounts to a 
relative systematic error on the energy of $0.02\%$ and is comparable to 
our statistical error bars. For a detailed review of this approach we 
refer the reader to Ref.~\cite{Assaad2008}. For the implementation we have 
used the ALF library \cite{ALF2017}. To carry out the analytic 
continuation, we have used the stochastic MaxEnt implementation 
\cite{Sandvik98,Beach04a} of the ALF library \cite{ALF2017}.

\begin{figure}[t!]
\centering
\includegraphics[width=0.97\columnwidth]{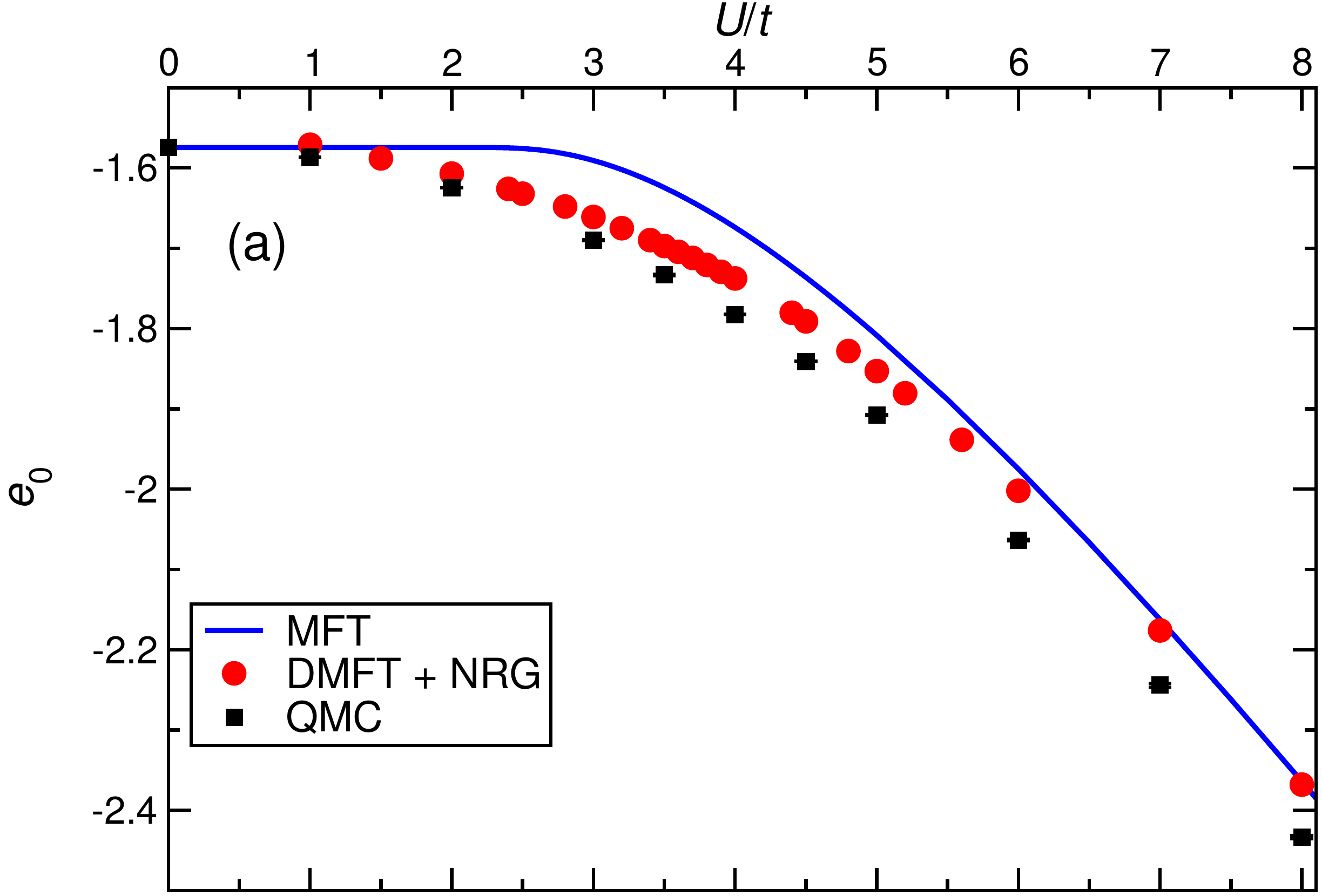} \\
\includegraphics[width=0.97\columnwidth]{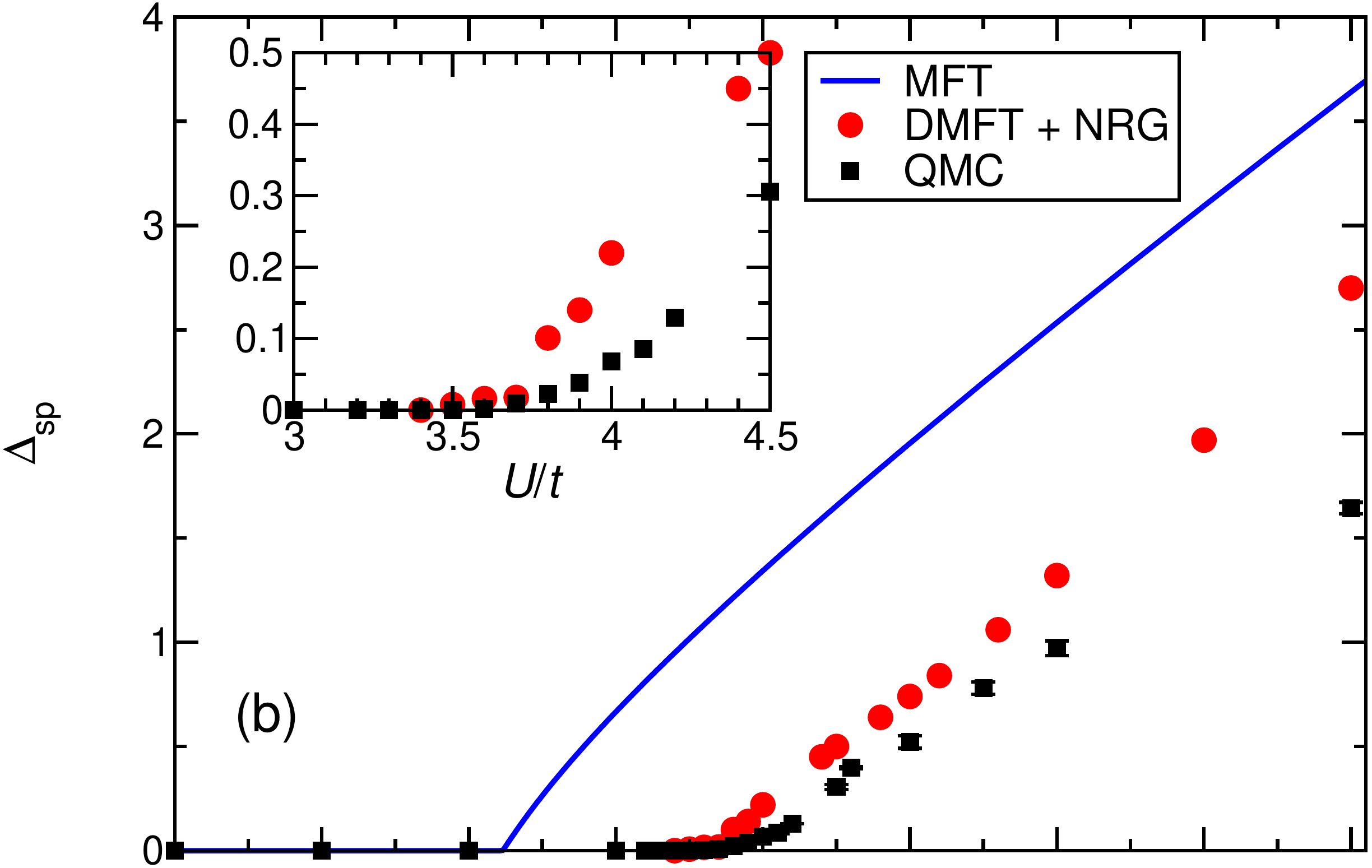} \\
\includegraphics[width=0.97\columnwidth]{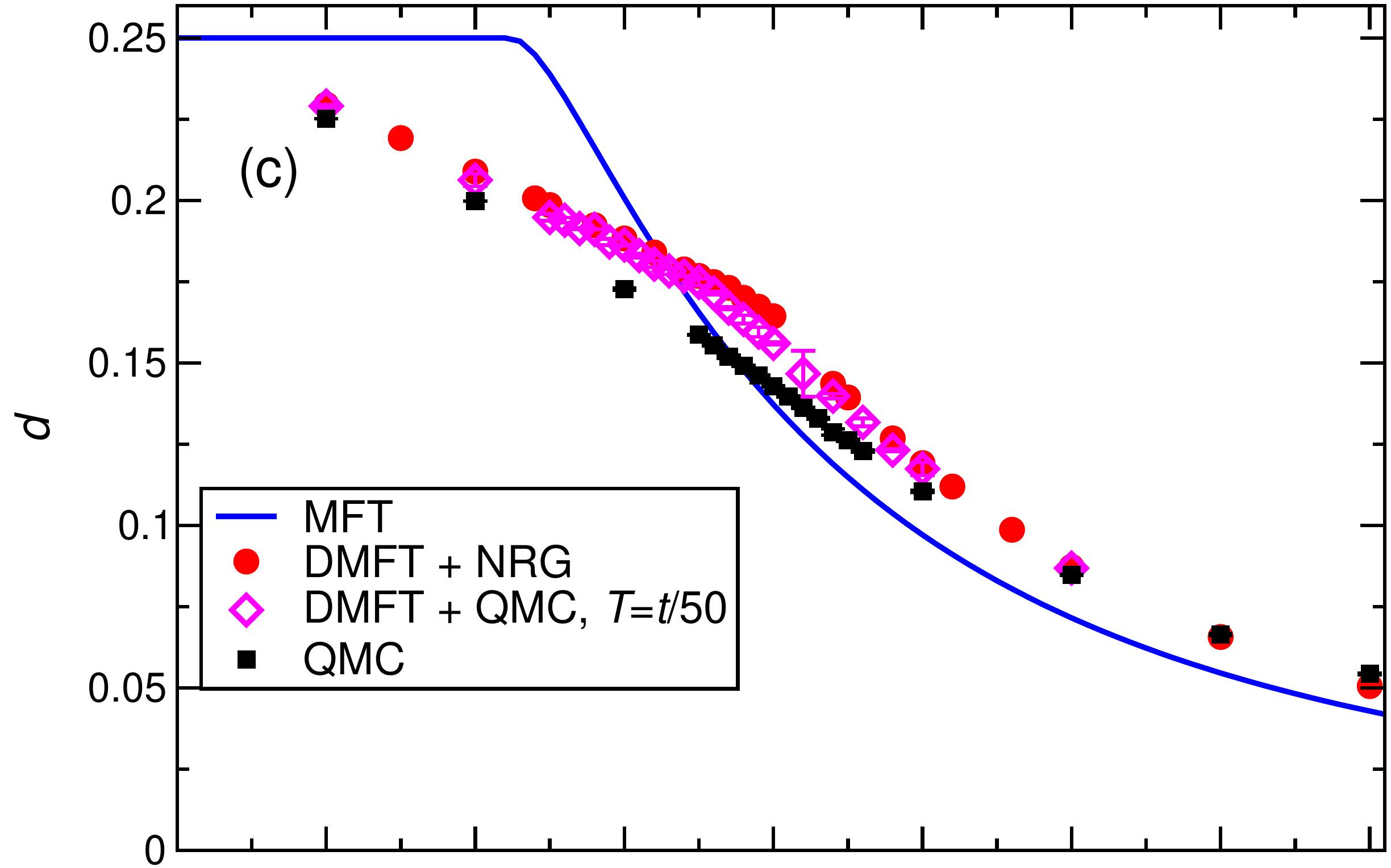} \\
\includegraphics[width=0.97\columnwidth]{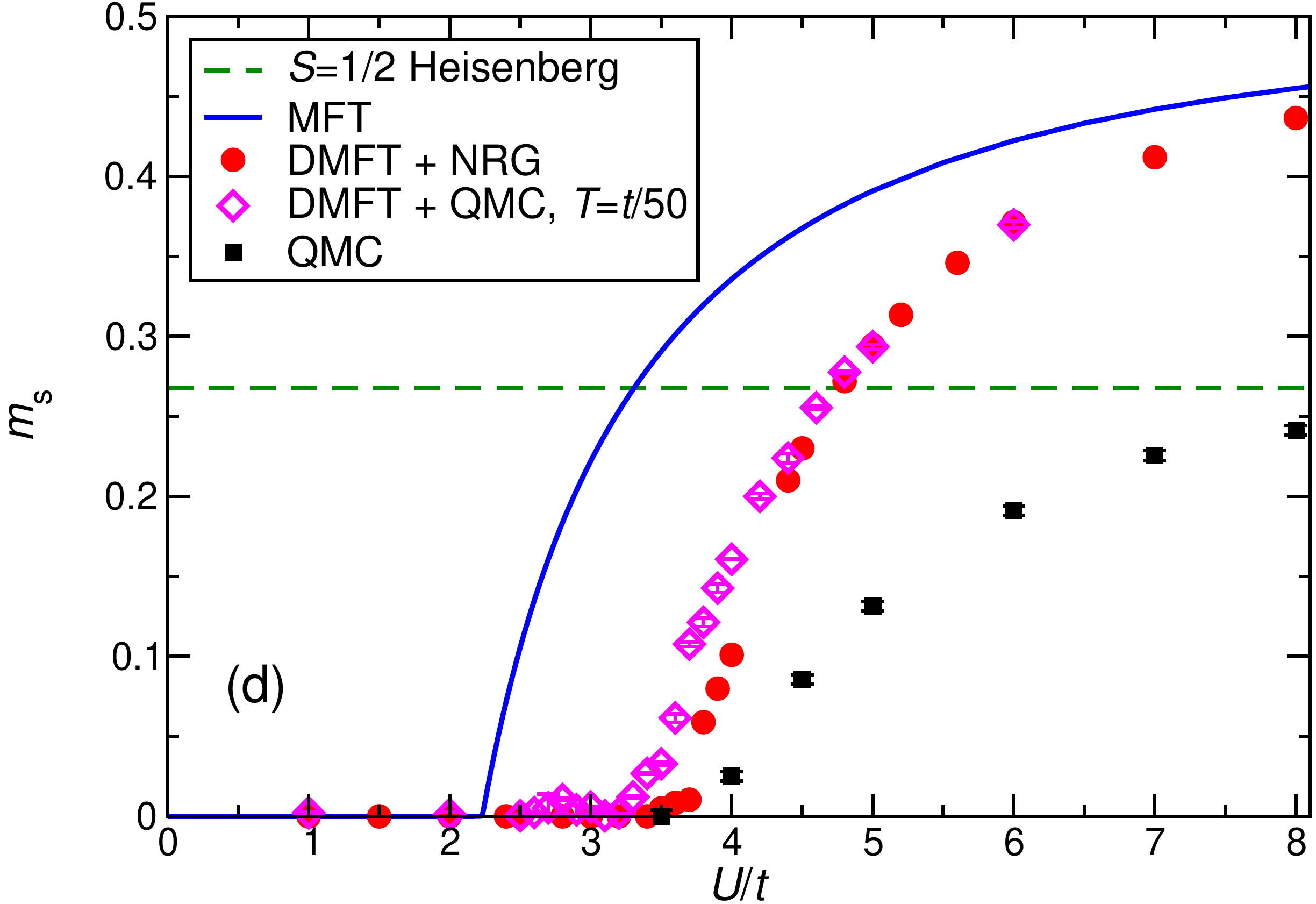}
\caption{Ground-state energy per site $e_0$ (a),
single-particle gap $\Delta_{\rm sp}$ (b),
double occupancy $d$ (c), and staggered
magnetization $m_{\rm s}$ (d) of an infinite honeycomb lattice.
\label{fig:Inf}
}
\vspace*{-10mm}
\end{figure}

\section{Static properties}

\label{eq:secInfStatic}

Figure \ref{fig:Inf} presents a comparison of static quantities that is 
similar in spirit to the QMC versus MFT comparison of Ref.~\cite{Feldner2010,*FeldnerE} 
for an $N=162$ site system subject to periodic boundary conditions, except 
that it is now for an infinite system and includes the single-site DMFT in 
the comparison. QMC results are partially taken from 
Ref.~\cite{Assaad2013}, supplemented by additional data points in order to 
cover a broader range and new data for the energy (not considered in 
Ref.~\cite{Assaad2013}). For our purposes, a system with $18 \times 18$ 
primitive cells, {\it i.e.}, $N=648$ sites can usually be considered as 
representative of the thermodynamic limit. For the DMFT, we focus on 
results obtained with a fast NRG impurity solver, but include results 
obtained from a slower QMC impurity solver for two quantities in 
Fig.~\ref{fig:Inf} in order to assess the effect of the different 
approximations in the impurity solver on top of the DMFT approximation.

The fact that total energies per site $e_0$ agree well 
(Fig.~\ref{fig:Inf}(a)) is a prerequisite for also more sensitive 
quantities to be in good agreement. Still, one can already see that the 
inclusion of charge fluctuations in DMFT improves over the static MFT, in 
particular for small to intermediate values of $U$ (for large values of 
$U$, DMFT approaches again the static MFT result). In addition, one may 
observe that the MFT result for $e_0$ based on the Hamiltonian 
(\ref{eq:HMF1}--\ref{eq:HMF3}) starts to deviate from its $U=0$ value only 
for $U> U_{c,{\rm MFT}}$ (actually, $U_{c,{\rm MFT}} \approx 2.23\,t$ 
\cite{Sorella1992} is more clearly identified in other quantities to be 
discussed below). The same behavior is observed also in other quantities 
and can be traced to the densities being pinned at $\langle n_{i,\sigma} 
\rangle =\frac{1}{2}$ for $U<U_{c,{\rm MFT}}$ on the infinite honeycomb 
lattice.

Next we turn to the staggered magnetization $m_{\rm s}$ shown in 
Fig.~\ref{fig:Inf}(d). The QMC results shown here differ from those of 
Ref.~\cite{Assaad2013} in so far as they were computed directly from the 
$\ve{Q}=(0,0)$ spin structure factor:
\begin{eqnarray}
	S_{\textrm{AF}} & =& \frac{1}{L^2} \sum_\alpha \sum_{\ve{r}} \langle \ve{S}_{\alpha}(\ve{r}) \cdot \ve{S}_{\alpha} \rangle, \nonumber \\
              m_{\rm s} & =& \sqrt{ \frac{S_{\textrm{AF}}} {N} },
\label{eq:S}
\end{eqnarray}
rather than with the aid of a pinning field. The numerical accuracy
of NRG being limited by the logarithmic discretization of the frequency axis,
values of $m_{\rm s} \lesssim 0.01$ can be considered to be zero
within DMFT+NRG. Consequently, in the DMFT+NRG data, we observe a rapid
increase of $m_{\rm s}$ around $U/t\approx 3.7$, signaling the onset of
magnetism. Thus, we find that the inclusion of charge fluctuations in
the DMFT shifts the transition from $U_{c,{\rm MFT}}/t \approx 2.23$ \cite{Sorella1992}
much closer to the ``exact'' QMC result $U_c/t \approx 3.78$ \cite{Assaad2013}.
Figure \ref{fig:Inf}(d) also shows data obtained from DMFT+QMC. QMC differs from
NRG in that it works on the imaginary frequency axis and at finite temperature
(the present data has been obtained at $T=t/50$). Thus, we can compare the effect
in particular of finite temperature within QMC and the effect of discretization and broadening
of the real-frequency spectral functions in NRG. First, we observe overall good
agreement with the biggest differences arising in the critical region. Since it is
difficult to say which DMFT variant is more reliable, we conclude from the comparison
that the critical point may shift down to $U_c/t\approx 3.5$ within DMFT. Despite this
uncertainty within DMFT, the value obtained by DMFT is in any case much closer to the
``exact'' QMC result than static mean-field theory.
This good correspondence extends even a bit into the magnetic phase owing to the
fact that the mean-field critical exponent $\beta=1$ for the staggered magnetization
(also valid for DMFT) is close to the true value $\beta = 0.8$ \cite{Assaad2013},
{\it i.e.}, the main difference just beyond the critical point seems to be a larger prefactor
for DMFT. This is also evident deep inside the magnetic phase. Again,
since DMFT is a mean-field theory, it yields $\lim_{U/t\to\infty} m_{\rm s} = 1/2$.
On the other hand, for $U \gg t$, the half-filled Hubbard model maps onto the spin-1/2
Heisenberg model on the same lattice. The staggered magnetization of the spin-1/2
Heisenberg model is reduced by quantum fluctuations and has been intensively studied
for the honeycomb lattice by a broad range of methods
\cite{Reger1989,Weihong1991,Oitmaa1992,Krueger2001,Richter2004,Castro2006,Jiang2008}.
Figure \ref{fig:Inf}(d) shows the estimate $m_{\rm s} = 0.2677(6)$
\cite{Castro2006} for the spin-1/2 Heisenberg model as a dashed horizontal line.
The QMC results for the full Hubbard model remain indeed systematically below
this line and might approach it asymptotically in the large-$U$ limit.

Now we briefly comment on the single-particle gap that is one half the 
charge gap, $\Delta_{\rm sp} = (E_{N-1} - 2\, E_{N} + E_{N+1})/2$, where 
$E_n$ is the ground-state energy in the sector with $n$ electrons. The 
single-particle gap $\Delta_{\rm sp}$ is shown in Fig.~\ref{fig:Inf}(b); 
it opens in the magnetic phase and thus exhibits similar behavior as the 
staggered magnetization. This is particularly evident in the MFT theory 
where $\Delta_{\rm sp}$ and $m_{\rm s}$ are directly related by 
Eq.~(\ref{eq:MFTspGap}). The DMFT+NRG result in Fig.~\ref{fig:Inf}(b) is 
remarkably close to the ``exact'' lattice QMC and just overestimates the 
gap a bit. The simple static MFT is again less accurate, as is expected in 
view of it underestimating the critical value $U_c$.

Finally, Fig.~\ref{fig:Inf}(c) shows the double occupancy
\begin{equation}
d = \frac{1}{N} \sum_i \left\langle n_{i,\uparrow} \, n_{i,\downarrow} \right\rangle \, .
\label{eq:DoubleOcc}
\end{equation}
The double occupancy has the advantage that it is related to the magnetic 
behavior of the system while being more easily accessible by QMC than spin 
expectation values. The actual QMC data shown in Fig.~\ref{fig:Inf}(c) is 
for $N=648$, but finite-size effects are negligible. We observe first that 
all three methods yield quantitatively similar results. The MFT transition 
$U_{c,{\rm MFT}}$ can be detected as the point where the double occupancy 
starts to fall below the $U=0$ value $d=1/4$, but MFT misses the emergence 
of a local moment in the paramagnetic phase as signaled by a drop in $d$. 
This reduction of the double occupancy and the resulting emergence of a 
local moment are much better reproduced by DMFT that yields results that 
are significantly closer to the ``exact'' QMC result for the Hubbard model 
than plain MFT, {\it i.e.}, inclusion of local charge fluctuations yields 
a substantial quantitative improvement. Figure \ref{fig:Inf}(c) compares 
again the NRG and QMC variants of DMFT. In this case, the difference 
between the two impurity solvers is found to be very small. However, there 
is no clear signal of the magnetic transition in $d$, neither in the DMFT 
nor in the lattice QMC results, {\it i.e.}, the double occupancy is not 
very useful for locating the transition point.

Overall, we find that DMFT improves static properties in the semi-metallic 
phase by including local charge fluctuations beyond static MFT. 
Specifically, these fluctuations affect the ground-state energy 
(Fig.~\ref{fig:Inf}(a)) and double occupancy (Fig.~\ref{fig:Inf}(c)), and 
shift these quantities close to the ``exact'' QMC results while within MFT 
these quantities remain pinned at their non-interacting $U=0$ values 
throughout the paramagnetic semi-metallic phase. Even the estimate for the 
critical $U_c$ turns out to be remarkably accurate within DMFT. Just 
deeper in the magnetic phase one observes larger deviations between DMFT 
and QMC. In particular, DMFT fails to account for the reduction of the 
magnetic moment at large $U/t$ by quantum fluctuations (see 
Fig.~\ref{fig:Inf}(d)) that would require a proper treatment of their 
spatial nature. Still, DMFT, in particular in the DMFT+NRG incarnation 
appears to be a remarkably accurate tool for describing the semi-metallic 
phase up to the region around $U_c$.

\section{Spectral functions}

\label{sec:SpecFns}

\begin{figure}[t!]
\centering
\includegraphics[width=0.95\columnwidth]{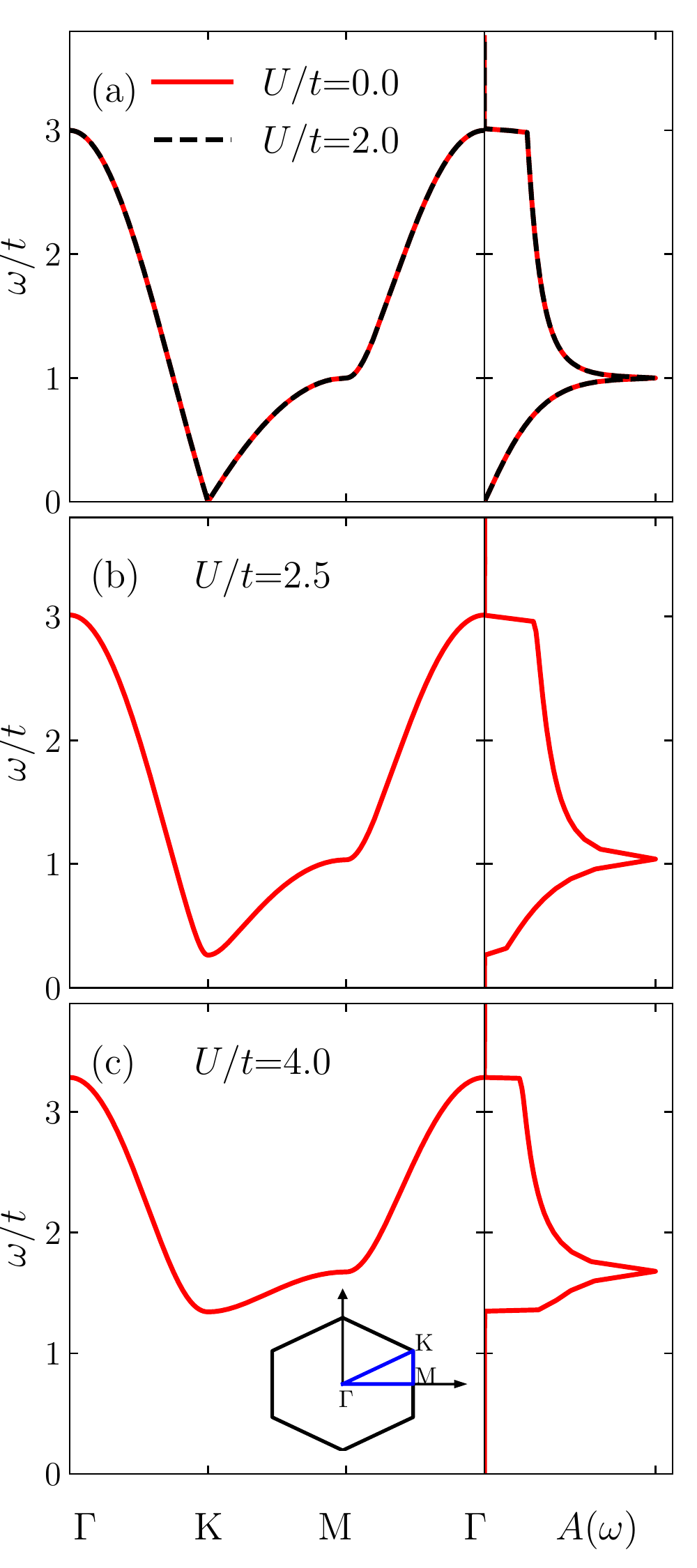}
\caption{Mean-field result for the single-particle dispersion (left) and
local spectral function $A(\omega)$ for $U/t=0$, $2$ (a), $2.5$ (b), and $4$ (c).
\label{fig:AkomegaMFT}
}
\end{figure}

\subsection{Static mean-field theory (MFT)}

First, we discuss the MFT results for the single-particle spectral functions.
Within MFT, the retarded Green's function reads: 
\begin{equation}
\label{Green_MF.eq}
G^{\text{ret}}_{\sigma} (\ve{k}, \omega ) = \frac{1} { \omega + i 0^{+} - \left( \tau_x \text{Re}z(\ve{k}) + \tau_y \text{Im}z(\ve{k}) - U\,\sigma \,m_{\rm s}\, \tau_z \right)} 
\end{equation}
such that the spin-averaged single-particle spectral function becomes: 
\begin{eqnarray}
A(\ve{k},\omega) &=& - \text{Im} \sum_{\sigma} \text{Tr} G^{\text{ret}}_{\sigma} (\ve{k}, \omega ) \nonumber \\
&=& 2 \pi \left[ \delta ( E(\ve{k}) - \omega) + \delta ( E(\ve{k}) + \omega) \right] \, .
\label{eq:MFTspec}
\end{eqnarray}
Thus, within MFT the spectral functions consists of $\delta$-functions at
the single-particle energy $\pm E(\ve{k})$.
The left column of Fig.~\ref{fig:AkomegaMFT} shows the mean-field
single-particle dispersion Eq.~(\ref{eq:MFTdisp}).
The spectra are reflection symmetric
$A(\ve{k},\omega) = A(\ve{k},-\omega)$ thanks to the particle-hole symmetry
\cite{Scalettar16,Wakabayashi13} and the two sites in the primitive cell of the one-band Hubbard model on the honeycomb lattice. Therefore, here and below we only show positive frequencies $\omega \ge 0$.

Since the matrix elements of the spin-averaged spectral function are 
constant, see Eq.~(\ref{eq:MFTspec}), the local density of states (or 
local spectral function) $A(\omega)$ is obtained by simple 
$\ve{k}$-integration of the MFT dispersion. The result is shown by the 
right column of Fig.~\ref{fig:AkomegaMFT}.

We observe in panel Fig.~\ref{fig:AkomegaMFT}(a) that at the mean-field 
level and in the semi-metallic phase $U < U_{c,{\rm MFT}}$, the Coulomb 
interaction $U$ has no effect on these observables since the mean field 
vanishes identically (compare a similar remark made for static observables 
in Sec.~\ref{eq:secInfStatic}). Consequently, we recover both the 
well-known dispersion and density of states of non-interacting 
tight-binding electrons on the honeycomb lattice, see, e.g., 
Refs.~\cite{RevModPhys81,CastroNeto2012,Wakabayashi13}. On the other hand, 
for $U > U_{c,{\rm MFT}}$, one observes first the opening of a gap at the 
K point (compare the examples for $U/t = 2.5$ and $4$ in 
Fig.~\ref{fig:AkomegaMFT}(b,c)), an increase of the total bandwidth, and a 
shift of the sharp peak in the middle of the spectra to higher values of 
the frequency $\omega$, in accordance with Eq.~(\ref{eq:MFTdisp}).

\subsection{Dynamical mean-field theory (DMFT)}

\begin{figure}[ht!]
\centering
\includegraphics[width=0.95\columnwidth]{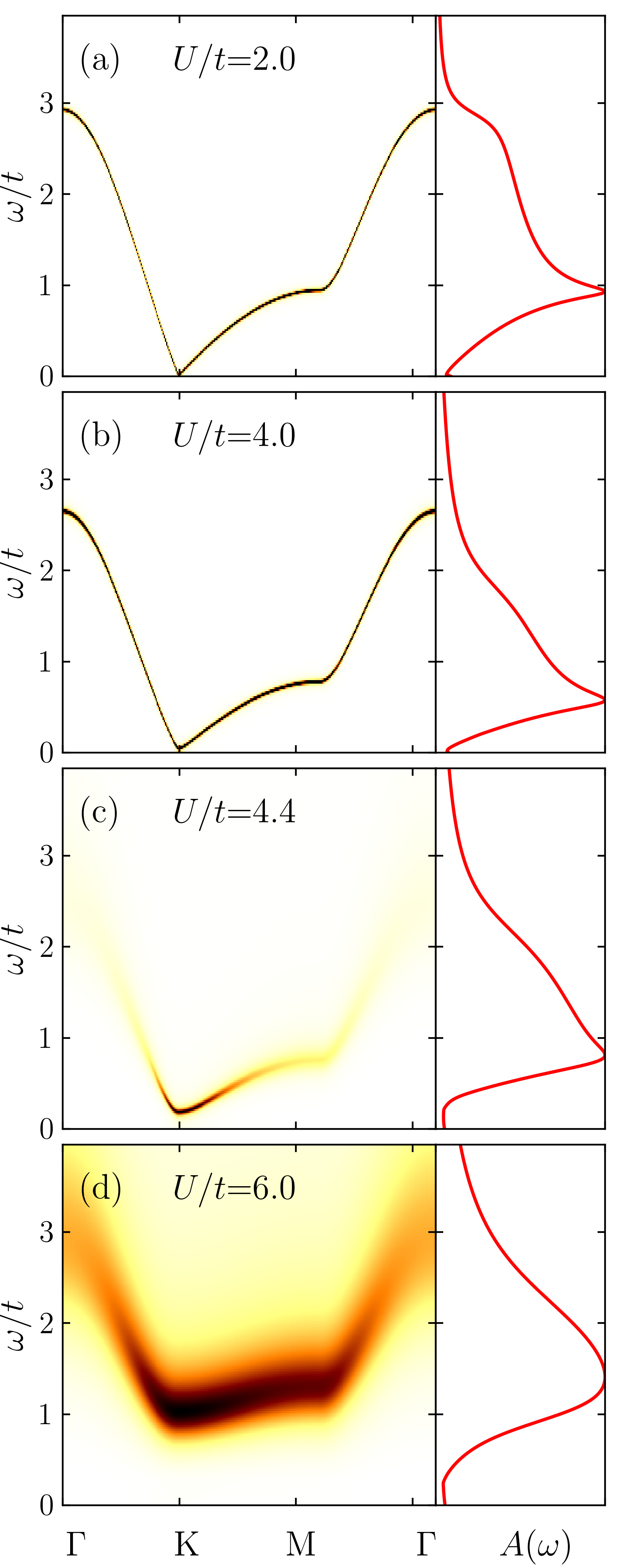}
\caption{DMFT result for the spectral function $A(\ve{k},\omega)$ (left)
and local spectral function $A(\omega)$
for $U/t=2$, $4$, $4.4$, and $6$ (top to bottom).
\label{fig:AkomegaDMFT}
}
\end{figure}

Figure \ref{fig:AkomegaDMFT} shows DMFT results obtained with the NRG 
impurity solver for the $\ve{k}$-resolved and local spectral function. In 
the left column of Fig.~\ref{fig:AkomegaDMFT}, we use a color coding to 
indicate the spectral weight of $A(\ve{k},\omega)$. Although the 
non-vanishing self energy $\mathbf{\Sigma}$ does modify the spectral 
functions also in the semi-metallic phase $0 < U < U_c$, this effect 
remains small. This is illustrated by the case $U/t=2$ in 
Fig.~\ref{fig:AkomegaDMFT}(a) that is very similar to the $U=0$ case, see 
Fig.~\ref{fig:AkomegaMFT}(a). The main difference is a small reduction in 
bandwidth (see $\omega/t \lesssim 3$) although we recall that the 
resolution of NRG at these high energies is limited.

The case $U=4\,t$ shown in Fig.~\ref{fig:AkomegaDMFT}(b) is already in the 
antiferromagnetic phase. Consequently, there should be a gap in the 
spectrum (compare also Fig.~\ref{fig:Inf}(b)), but it is too small to be 
visible in Fig.~\ref{fig:AkomegaDMFT}(b). In DMFT, the magnetization and 
the correlations inherent in the system are still comparably small for 
$U/t=4$. Thus, the gap due to antiferromagnetic order is small. 
Furthermore the broadening due to an imaginary part of the self energy is 
small; the lifetime of the quasiparticles is very long. However, upon 
increasing the interaction strength to $U/t=4.4$ 
(Fig.~\ref{fig:AkomegaDMFT}(c)), the gap as well as the broadening of the 
quasiparticle bands become visible. For $U/t=6$ 
(Fig.~\ref{fig:AkomegaDMFT}(d)), the lifetime of the particle becomes 
short and the bands are strongly broadened due to the self energy. 
Furthermore, because of Hubbard satellites at $E=\pm U/2$ the bandwidth 
becomes enhanced.

From the symmetry point of view, the DMFT approximation explicitly breaks 
the SU(2) spin symmetry. This explicit versus spontaneous symmetry 
breaking has for consequence that spatial spin fluctuations encoded in the 
Goldstone modes are absent. As such the DMFT spectral function should be 
understood in terms of a particle propagating in a frozen 
antiferromagnetic environment, as in the static mean-field approximation. 
In fact and from the weak to intermediate coupling limit, the DMFT results 
presented in Fig.~\ref{fig:AkomegaDMFT} exhibit a spectral function very 
similar to the mean-field approximation albeit with a broadening due to 
the imaginary part of the self energy that becomes significant for 
$U/t=4.4$ and $6$, compare Fig.~\ref{fig:AkomegaDMFT}(c,d).

\subsection{Lattice QMC}

\begin{figure}[h!]
\centering
\includegraphics[width=0.97\columnwidth]{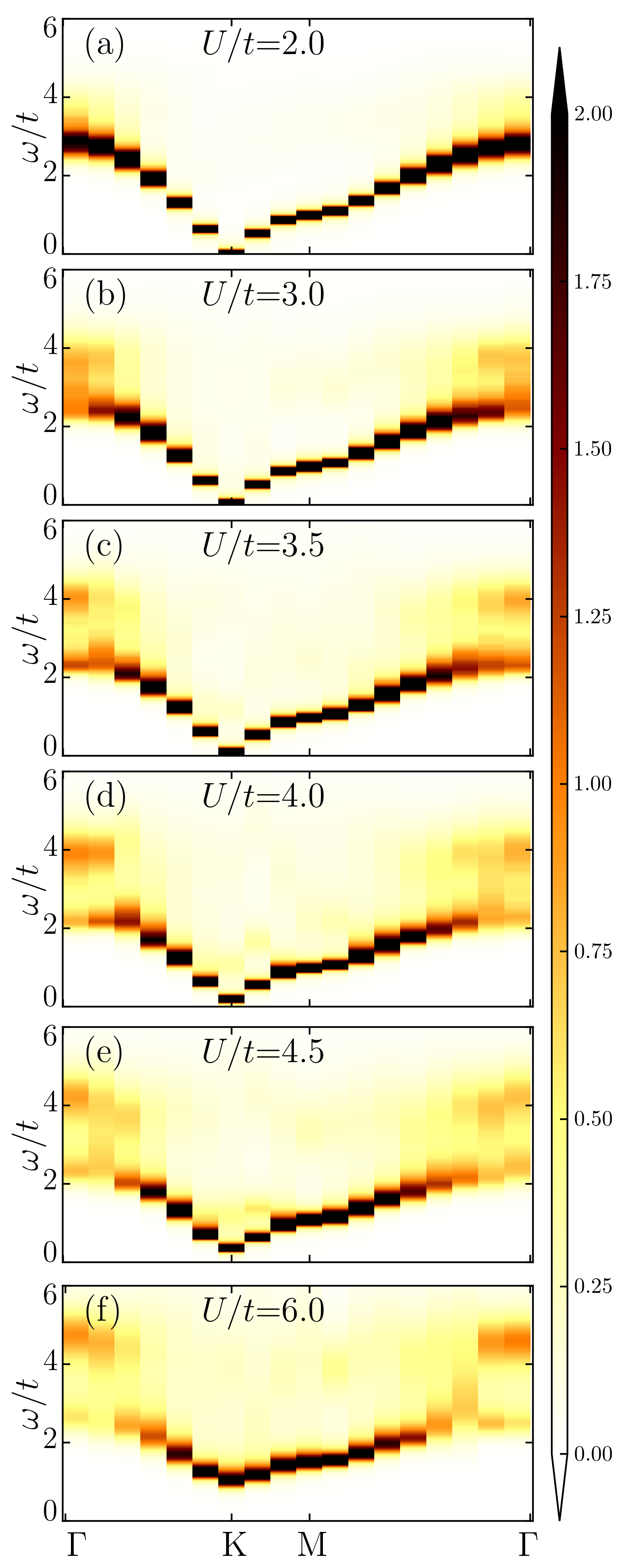}
\caption{QMC result for the spectral function $A(\ve{k},\omega)$
on a honeycomb lattice of $18 \times 18$ primitive cells.
\label{fig:AkomegaQMC}
}
\end{figure}

In the lattice QMC approach the SU(2) spin symmetry is spontaneously 
broken. As mentioned above this gives rise to collective spin-wave 
excitations (Goldstone modes) that, as we will see, have a big impact on 
the single-particle spectral function. Our results are plotted in 
Fig.~\ref{fig:AkomegaQMC} across the metal-insulator transition. In the 
weak-coupling limit, $U/t=2$, the data shown in 
Fig.~\ref{fig:AkomegaQMC}(a) agrees within numerical accuracy with the 
DMFT result of Fig.~\ref{fig:AkomegaDMFT}(a) and consequently also with 
the one from static MFT.

As appropriate for the Gross-Neveu transition at $U_c$, the velocity 
remains finite, and to a first approximation the opening of the gap 
follows the mean-field form. The mean-field approximation becomes exact at 
the upper critical dimension corresponding to $d=3$. By contrast, in two 
spatial dimensions the single-particle propagator acquires an anomalous 
dimension, and we would expect a branch cut instead of a pole at the 
critical point. Within the $\epsilon$-expansion around $d=3$ and at first 
order \cite{Herbut09a}, the fermion anomalous dimension is given by 
$\eta_f = 0.03$. This small value is consistent with the fact that we do 
not observe a broadening of the spectral function in the vicinity of the 
critical coupling (Fig.~\ref{fig:AkomegaQMC}(c,d)) and at the Dirac point 
K. We note that this is very similar to the order-disorder transition as 
realized by the Heisenberg model on a bilayer lattice. Here the anomalous 
dimension of the bosonic model is equally very small, such that even at 
the critical point we observe a sharp feature in the dynamical spin 
structure factor \cite{Lohofer15}.

Beyond the critical coupling $U_c/t \approx 3.78$ \cite{Assaad2013}, the 
data for the spectral function corresponds to the motion of a single hole 
in a quantum antiferromagnet. In conjunction with the cuprates, this 
problem has been extensively studied on the square lattice 
\cite{Preuss95,Martinez91,Brunner00b}. On the honeycomb lattice, the 
spectral function shows two prominent features that are especially visible 
at the $\Gamma$ point starting from $U/t=3.5$ 
(Fig.~\ref{fig:AkomegaQMC}(c)).
First, there is an \textit{incoherent} high-energy feature
that shifts to higher energies with increasing $U/t$.
The second low-energy feature for 
$\omega/t \lesssim 2$ is much sharper. We therefore interpret it as a 
coherent quasiparticle band, the width of which decreases with increasing 
$U/t$. As in the $t$-$J$ model on the square lattice 
\cite{Martinez91,Brunner00b}, one expects the bandwidth of this coherent 
band to scale as the magnetic scale $J \simeq t^2/U$ reflecting the fact 
that hole motion scrambles the spin background and that the healing 
procedure can only occur on a time scale set by $J$. We will hence adopt 
the same terminology as on the square lattice and refer to the coherent 
feature as the ``spin polaron''.

The generic form of the zero-temperature spectral function in the Lehmann 
representation reads $A_{\sigma} (\ve{k}, \omega) = \pi \sum_n | \langle n 
| c^{\dagger}_{\ve{k},\sigma} | 0 \rangle |^2 \, \delta (E_n - E_0 - 
\omega) + \pi \sum_n | \langle n | c_{\ve{k},\sigma} | 0 \rangle |^2\, 
\delta (E_0 - E_n - \omega)$. Here $ H | n \rangle = E_n |n \rangle $ and 
the sum rule $\int {\rm d} \omega\, A_{\sigma} (\ve{k}, \omega) = \pi$ 
holds. Hence both the energy spectrum and the matrix elements are required 
for a full understanding of the spectral function. In particular the 
support of the spectral function is given by the energy spectrum and the 
distribution of weight by the matrix elements. At our largest coupling, 
$U/t = 6$, it is apparent from Fig.~\ref{fig:AkomegaQMC}(f) that at the 
$\Gamma$ point the dominant weight is in the incoherent high-energy feature
and that 
this spectral weight is transferred to the coherent spin-polaron band upon 
approaching the M or K point. This rather abrupt transfer of spectral 
weight is referred to as \textit{waterfall} in the high-$T_c$ literature 
and has been observed in simulations of the Hubbard model on the square 
lattice \cite{Preuss95,Moritz10} as well as experimentally in 
photoemission studies of the cuprates \cite{Graf07}.

\section{Conclusions and discussion}

\label{sec:Conclusions}

We have performed a comparative investigation of the one-band Hubbard 
model on the honeycomb lattice, using static mean-field theory (MFT),
dynamical mean-field theory (DMFT), and ``exact'' 
quantum Monte Carlo (QMC) simulations on the lattice. All three methods 
yield a semi-metallic Dirac phase and an antiferromagnetic insulator. The 
critical point in MFT $U_{c,{\rm MFT}} \approx 2.23\,t$ \cite{Sorella1992} 
is significantly below the exact location $U_c/t \approx 3.78$ 
\cite{Assaad2013}. Our first finding is that the single-site DMFT yields a 
very good approximation to this value, namely $3.5 \lesssim U_c/t \lesssim 
3.7$ and is thus competitive in accuracy with more sophisticated 
generalizations of DMFT \cite{Wu2014,Hirschmeier2018}. In this respect, an 
accurate treatment of the effective impurity problem thus appears to be 
more important than going to big cluster sizes.

Within static MFT, all quantities are independent of $U$ for $U < U_c$ 
owing to the vanishing mean field. This is improved by DMFT, yielding in 
particular more accurate values of the ground-state energy and double 
occupancy. All three methods find qualitatively similar spectral functions 
in the semi-metallic phase with a sharp and gapless quasiparticle. The 
main improvement by DMFT in this case is a broader range in $U$ that is 
accessible owing to the better estimate for $U_c$. Overall, we find that 
single-site DMFT provides a remarkably accurate description of the weakly 
correlated semi-metallic phase at a low computational cost, in particular 
when the numerical renormalization group (NRG) \cite{Wilson1975,BCT08} is 
used as impurity solver.

Both simple MFT and DMFT yield mean-field critical behavior and are thus not expected to provide
quantitatively accurate results close to $U_c$ and in particular for the critical exponents although
the actual values for the relevant Gross-Neveu transition are quite close to the mean-field
values \cite{Assaad2013,Herbut09a}. For large values of $U$ deep inside the antiferromagnetic phase,
DMFT reduces again to static MFT and misses in particular the non-local spin fluctuations.
Thus, the staggered magnetization $m_{\rm s}$ tends to $1/2$ for $U \to \infty$ both within static MFT and DMFT,
{\it i.e.}, both methods fail to reproduce the reduction of the ordered moment at large $U$
by quantum fluctuations. For the same
reasons, DMFT and in particular MFT overestimate the single-particle gap that is induced by the magnetic
order in the magnetic phase.

As a first perspective for further work, we mention applications to 
magnetism induced at zig-zag edges of graphene-type nanostructures 
\cite{Yazyev2010,Wakabayashi13,Fujita96,Wakabayashi98,FernandezRossier2007,Bhowmick2008,Jiang08,Viana-Gomes2009,Feldner2010,*FeldnerE,Feldner2011,Roy2014,Valli2016,PRB.96.115155}. 
Previous studies \cite{Feldner2010,*FeldnerE,Feldner2011} observed that simple MFT is remarkably successful in describing at 
least some aspects of this phenomenon in the weakly correlated regime. In 
particular, the local spectral functions for nanoribbons turned out to be 
remarkably accurate in MFT \cite{Feldner2011}. The main shortcoming of MFT 
is that it underestimates the bulk critical value of $U_c$, thus limiting 
the range of $U$ where MFT applies. It is straightforward to generalize
the single-site DMFT employed in the present work to real-space nanostructures
in the same way as static MFT. Since our single-site DMFT yields a much better 
estimate for $U_c$, we speculate that a real-space variant will also 
further improve the description of edge-state magnetism beyond static MFT,
at least in the weakly correlated regime relevant to graphene, despite the
shortcomings of the single-site DMFT in the magnetic phase.

One of the biggest challenges in realistic DMFT-based calculations is to 
include non-local correlations \cite{Vollhardt2017}. We believe that this 
work provides a non-trivial benchmark to further test 
various schemes aimed at including non-local fluctuations around the DMFT 
solution. This includes dual fermions \cite{Rubtsov08}, the dynamical 
vertex approximation \cite{Toschi07}, as well as extended DMFT 
\cite{Smith00}. On the other hand one can start with implementations of 
the functional renormalization group \cite{Metzner12} approach that 
captures spatial correlations but neglects temporal ones. Irrespective of 
the starting point, the proposed benchmark is highly non-trivial since the 
critical point is Lorentz invariant such that long-wave-length 
fluctuations in space and time are identical.

\begin{acknowledgments}

This work was supported by the Deutsche Forschungsgemeinschaft (DFG) under 
grants FOR1807 and RA 2990/1-1, by the ANR project J2D (ANR-15-CE24-0017), 
the Ministry of Education and Training of the Socialist Republic of 
Vietnam via a 911 fellowship, the Paris//Seine excellence initiative, and 
by JSPS KAKENHI Grants No.\ 18K03511 and No.\ 18H04316 (JPhysics). The 
authors gratefully acknowledge the Gauss Centre for Supercomputing e.V.\ 
(www.gauss-centre.eu) for funding this project by providing computing time 
on the GCS Supercomputer SuperMUC-NG at Leibniz Supercomputing Centre 
(www.lrz.de). The DMFT simulations were performed on the ``Hokusai'' 
supercomputer in RIKEN and the supercomputer of the Institute for Solid 
State Physics (ISSP) in Japan. RP thanks the Universit\'e de 
Cergy-Pontoise and their Institute for Advanced Studies for hospitality 
during a research visit.

\end{acknowledgments}

\appendix*

\section{Structure of the self energy}

\label{app:SelfEnergy}

The symmetry differences between the QMC simulations and the DMFT 
approximation become evident when considering the self energy. The 
QMC simulations possess the full symmetry of the Hubbard model: SU(2) spin rotation, inversion, time reversal, as well as particle-hole symmetries.

Particle-hole symmetry, $\hat{P}$, is an anti-linear  transformation that maps: 
\begin{eqnarray}
& & 
	\hat{P}^{-1}   \alpha 
 \begin{pmatrix}
c_{A,\sigma}(\ve{k}) \\ c_{B,\sigma}(\ve{k})
\end{pmatrix} 
\hat{P}^{-1}  =  
\overline{\alpha}
 \begin{pmatrix}
c^{\dagger}_{A,-\sigma}(\ve{k}) \\ -c^{\dagger}_{B,-\sigma}(\ve{k})
\end{pmatrix} 
\nonumber \\ 
& & =  \overline{\alpha} \tau_z 
\begin{pmatrix}
c^{\dagger}_{A,-\sigma}(\ve{k}) \\ c^{\dagger}_{B,-\sigma}(\ve{k})
\end{pmatrix}.
\end{eqnarray}
Due to the SU(2) spin symmetry, the single-particle Green's function matrix is spin independent and satisfies the symmetry property:
\begin{equation}
    \ve{G}^{\text{ret}}  (\ve{k}, \omega )  =  -  \tau_z \, \ve{G}^{\text{ret}}  (\ve{k}, - \omega) \,  \tau_z.
\end{equation}

Inversion 
symmetry amounts to:
\begin{equation}
\hat{I}^{-1}
\begin{pmatrix}
c_{A,\sigma}(\ve{k}) \\ c_{B,\sigma}(\ve{k})
\end{pmatrix} \hat{I}
=
\begin{pmatrix}
c_{B,\sigma}(-\ve{k}) \\ c_{A,\sigma}(-\ve{k})
\end{pmatrix} = 
\tau_x \, 
\begin{pmatrix}
c_{A,\sigma}(-\ve{k}) \\ c_{B,\sigma}(-\ve{k})
\end{pmatrix} 
\end{equation}
and as a consequence, 
\begin{equation}
	\ve{G}^{\text{ret}} (\ve{k}, \omega )
	 = \tau_x \, \ve{G}^{\text{ret}} (-\ve{k}, \omega ) \, \tau_x \,.
\end{equation}

Finally, time reversal symmetry reads: 
\begin{equation}
\hat{T}^{-1}
\alpha
\begin{pmatrix}
c_{A,\sigma}(\ve{k}) \\ c_{B,\sigma}(\ve{k})
\end{pmatrix} \hat{T}
= \overline{\alpha}  \sum_{s} (i \sigma_y)_{\sigma,s} 
\begin{pmatrix}
c_{A,s}(-\ve{k}) \\ c_{B,s}(-\ve{k})
\end{pmatrix} 
\end{equation}
leading to 
\begin{equation}
	\ve{G}^{\text{ret}} (\ve{k}, \omega )
	 =   \ve{G}^{\text{adv}} (-\ve{k}, \omega )  \, . 
\end{equation}
One will readily check that  the non-interacting Green's function of Eq.~(\ref{Green_MF.eq})  at $m_{\rm s}=0$ satisfies all the above properties.

Owing to the Dyson equation, the aforementioned symmetries carry over to the self-energy matrix that has to satisfy:
\begin{equation}
	\ve{\Sigma}(\ve{k},\omega) = \tau_x \, \ve{\Sigma}(-\ve{k},\omega) \, \tau_x \,,
\end{equation}
\begin{equation}
     \ve{\Sigma} (\ve{k}, \omega ) =   - \tau_z \,  \ve{\Sigma} (\ve{k}, -\omega ) \, \tau_z \,,
\end{equation}

Let us now parameterize the self energy as 
\begin{equation} 
  \ve{ \Sigma} (\ve{k},\omega) = \Sigma_x(\ve{k},\omega) \, \tau_x
       + \Sigma_y(\ve{k},\omega) \, \tau_y
       + \Sigma_z(\ve{k},\omega) \, \tau_z   
       + \Sigma_0(\ve{k},\omega) 
\end{equation}
where $\Sigma_\alpha(\ve{k},\omega)$ are scalar functions.    Inversion and particle-hole symmetry then implies that:
\begin{eqnarray}
	&& \Sigma_x(\ve{k},\omega) = \phantom{-} \Sigma_x(-\ve{k},\omega)  =  \phantom{-}  \Sigma_x(\ve{k},-\omega)    \nonumber \\
	&& \Sigma_y(\ve{k},\omega) = -                   \Sigma_y(-\ve{k},\omega)  =  \phantom{-}  \Sigma_y(\ve{k}, -\omega)    \nonumber \\
	&& \Sigma_z(\ve{k},\omega) = -\Sigma_z(-\ve{k},\omega)  = -\Sigma_z(\ve{k},-\omega)       \nonumber \\
	& & \Sigma_0(\ve{k},\omega) =  \phantom{-}  \Sigma_0(-\ve{k},\omega)  = -\Sigma_0(\ve{k},-\omega) \,  .
\end{eqnarray}

Generically, one sees that $\Sigma_0$ and $\Sigma_z$ are odd functions of frequency whereas $\Sigma_x$ and $\Sigma_y$ are even functions of frequency. 
The above greatly simplifies at time-reversal symmetric points in the Brillouin zone, $ \overline{\ve{k}} 
= \Gamma$, M.   Here the self energy reads: 
\begin{equation}
      \ve{\Sigma}(\overline{\ve{k}}, \omega)   =  \tau_x  \Sigma_x (\overline{\ve{k}}, \omega) +  \Sigma_0 (\overline{\ve{k}}, \omega). 
\end{equation}

This stands in strong contrast to the single-site
DMFT approximation where the self energy is spin dependent and diagonal in 
orbital space:
\begin{equation}
	\ve{\Sigma}^{\rm DMFT}_{\sigma} (\omega) =  \sigma \, \tau_z \, \Sigma^{\rm DMFT}( \omega).
\end{equation}
We note that the DMFT self energy satisfies particle-hole symmetry  but violates inversion as well as time reversal.

Returning to the QMC simulations, we have the following relation at the
$\Gamma$ point:
\begin{eqnarray}
 \text{Tr}  \, \ve{G}^{\text{ret}} (\Gamma,\omega) 
 &=& \frac{2 ( \omega - \Sigma_0(\Gamma,\omega) ) }{( \omega  + i0^{+}  - \Sigma_0(\Gamma,\omega) )^2  - (3 + \Sigma_x(\Gamma,\omega))^2 } \, .  \nonumber \\
 \end{eqnarray}

In contrast for the DMFT calculation we obtain: 
\begin{eqnarray}
 \text{Tr}  \, \ve{G}_{\sigma}^{\text{ret},DMFT} (\Gamma,\omega) 
 &=& \frac{2 \omega }{( \omega  + i0^{+})^2  - (\Sigma^{DMFT}(\Gamma,\omega) )^2  -  9 } \, .  \nonumber \\
\end{eqnarray}
 
Hence we can certainly compare the spectral functions, but
comparison of the self energy seems difficult.

\bibliography{graphene-dmft,fassaad,marcin}

\end{document}